\documentclass[sigconf,anonymous=false]{acmart}
\usepackage{url}  
\usepackage{graphicx}  
\usepackage{microtype}
\usepackage{booktabs} 
\usepackage[ruled,vlined,linesnumbered,resetcount]{algorithm2e}
\usepackage{arydshln}
\usepackage{subcaption}
\usepackage{flushend}
\usepackage{amsmath}
\usepackage{booktabs}
\usepackage{threeparttable}
\usepackage{multicol}
\usepackage{bbm}
\usepackage{amssymb}
\usepackage{pifont}
\usepackage[english]{babel}
\usepackage[font=small,labelfont=bf,tableposition=top]{caption}
\DeclareCaptionLabelFormat{andtable}{#1~#2  \&  \tablename~\thetable}
\setcopyright{rightsretained}

\acmDOI{10.475/123_4}

\acmISBN{123-4567-24-567/08/06}

\acmConference[WWW'19]{ACM  conference}{May 2019}{San Francisco}
\acmYear{2019}
\copyrightyear{2016}

\acmArticle{4}
\acmPrice{15.00}

\begin{document}
\title{Collaborative Similarity Embedding for Recommender Systems}
\author{Chih-Ming Chen}
\authornote{Social Networks and Human-Centered Computing,
Taiwan International Graduate Program,
Institute of Information Science, Academia Sinica, Taiwan}
\affiliation{%
  \institution{National Chengchi University}
  \city{Taipei, Taiwan}
}
\email{104761501@nccu.edu.tw}

\author{Chuan-Ju Wang}
\affiliation{%
  \institution{Academia Sinica}
  \city{Taipei, Taiwan}
}
\email{cjwang@citi.sinica.edu.tw}

\author{Ming-Feng Tsai}
\authornote{MOST Joint Research Center for AI Technology and All Vista Healthcare, Taiwan}
\affiliation{%
  \institution{National Chengchi University}
  \city{Taipei, Taiwan}
}
\email{mftsai@nccu.edu.tw}

\author{Yi-Hsuan Yang}
\affiliation{%
  \institution{Academia Sinica}
  \city{Taipei, Taiwan}
}
\email{yang@citi.sinica.edu.tw}





\begin{abstract}
We present collaborative similarity embedding (CSE), a unified framework that
exploits comprehensive collaborative relations available in a user-item
bipartite graph for representation learning and recommendation.
In the proposed framework, we differentiate two types of proximity relations:
direct proximity and $k$-th order neighborhood proximity.
While learning from the former exploits direct user-item associations
observable from the graph, learning from the latter makes use of implicit
associations such as user-user similarities and item-item similarities, which
can provide valuable information especially when the graph is sparse.
Moreover, for improving scalability and flexibility, we propose a sampling
technique that is specifically designed to capture the two types of proximity
relations.
Extensive experiments on eight benchmark datasets show that CSE yields
significantly better performance than state-of-the-art recommendation
methods.
\end{abstract}

%
%
%
%
%

\maketitle

\begin{figure*}
\centering
\begin{subfigure}[b]{0.6\textwidth}
    \includegraphics[width=\textwidth]{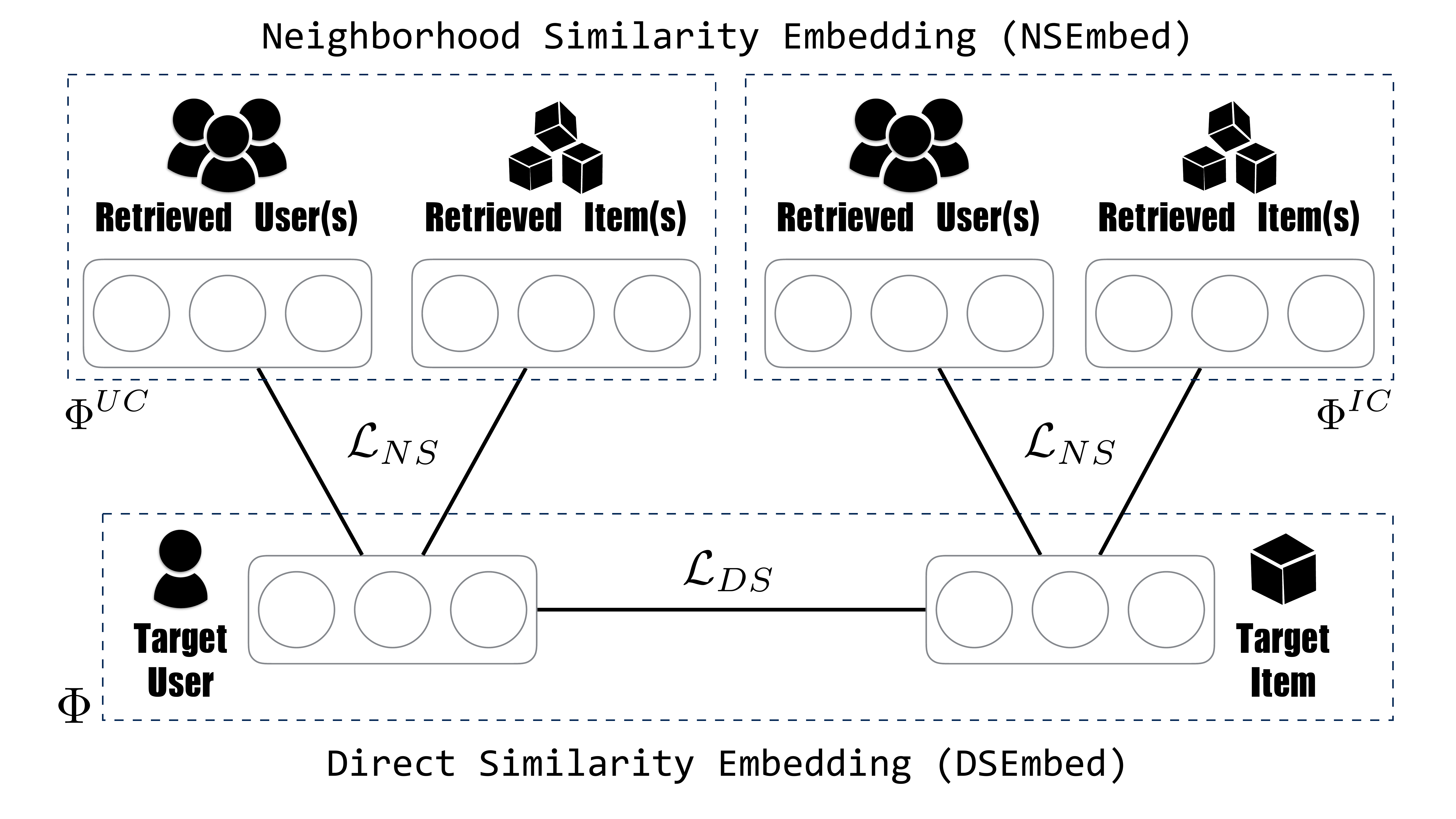}
    \caption{The bottom part depicts the direct similarity embedding module for
      user-item associations, whereas the upper left (right) part corresponds to
      modeling user-user (item-item, respectively) similarity with the
      neighborhood similarity embedding module.  An optimization step
    includes a sampled pair of a user and an item in DSEmbed and multiple high-order
  relation pairs in NSEmbed.}
\end{subfigure}
~
\begin{subfigure}[b]{0.38\textwidth}
    \includegraphics[width=\textwidth]{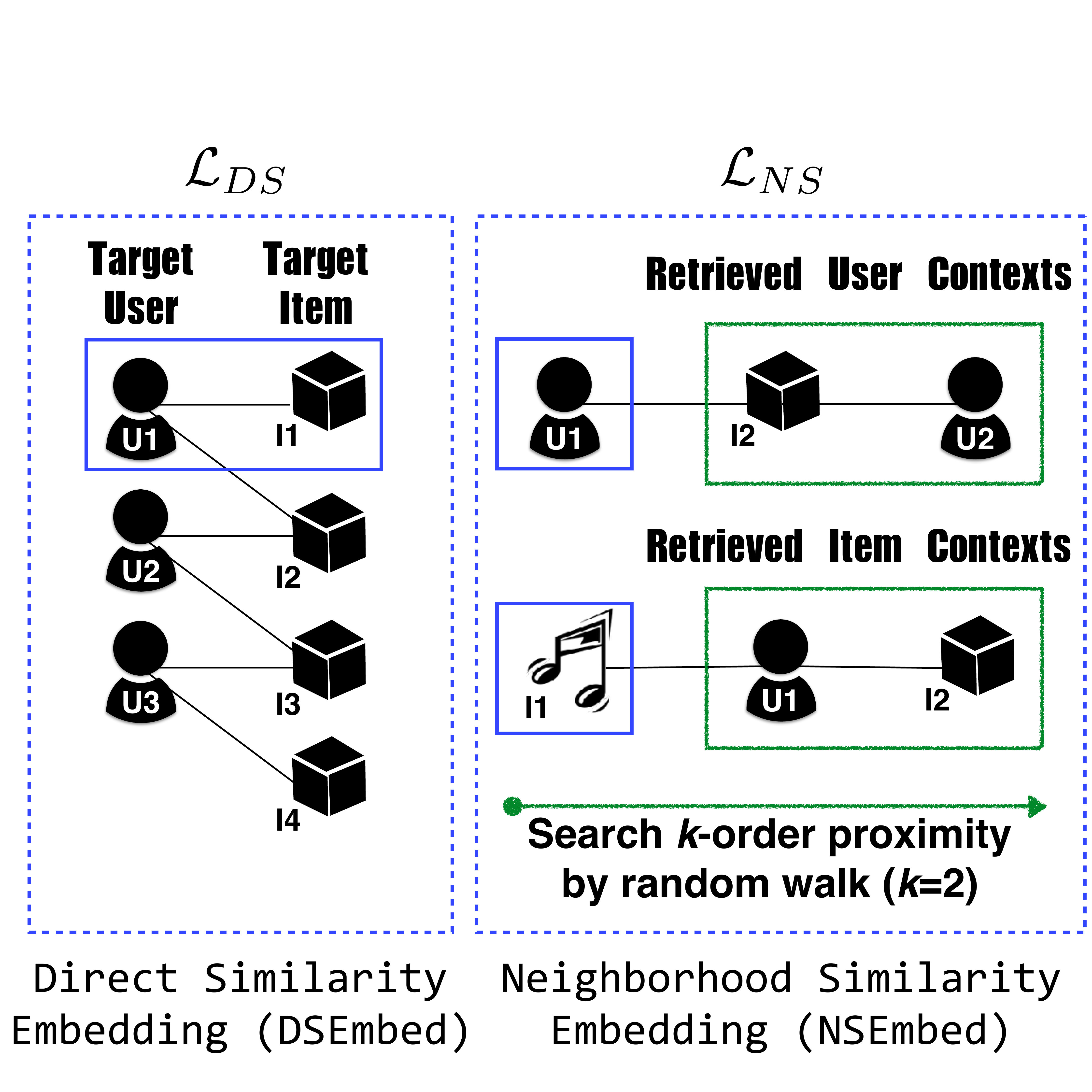}
    \caption{The left part shows that a target user-item pair
      (\texttt{U1}-\texttt{I1}) can be directly sampled from the observed edges
      for DSEmbed; the right part shows that for \texttt{U1} (or \texttt{I1}),
      a~$2$-step random walk is applied to obtain the contexts used in the
      NSEmbed module.}
\end{subfigure}
\caption{An overview of the proposed CSE framework}
\label{fig:overview}
\end{figure*}

\section{Introduction}

The task of recommender systems is to produce a list of recommendation results
that match user preferences given their past behavior.
Collaborative filtering (CF), a common yet powerful approach, generates
user recommendations by taking advantage of the collective wisdom from all
users~\cite{cacm}.
Many CF-based recommendation algorithms have been shown to work well across
various domains and been used in many real-world
applications~\cite{su2009survey}.

The core idea of model-based CF algorithms is to learn low-dimensional
representations of users and items from either explicit user-item associations
such as user-item ratings or implicit feedback such as playcounts and dwell
time.
This can be done by training a rating-based model with matrix completion to
learn from observed user-item associations (either explicit or implicit
feedback) to predict associations that are
unobserved~\cite{i2v,glslim,cml,wals,svd++,mf,cofactor,slim,e2r,nerm,pmf}.
In addition to this rating-based approach, ranking-based methods have been
proposed based on optimizing ranking loss; the ranking-based
methods~\cite{e2r,bpr,warp,kos,walkranker} have been found more suitable for
implicit feedback.
However, many existing model-based CF algorithms leverage only the user-item
associations available in a given user-item bipartite graph.
Thus, when the available user-item associations are sparse, these
algorithms may not work well.

It has been noted that it is possible to mine from a user-item bipartite graph
other types of collaborative relations, such as user-user similarities and
item-item similarities, since users and items can be indirectly connected in
the graph.
Moreover, by taking random walks on the graph, it is possible to exploit
higher-order proximity among users and items.
Using item-item similarities in the learning process has been firstly
studied by Liang \emph{et al.}~\cite{cofactor}, who propose to jointly
decompose the user-item interaction matrix and the item-item co-occurrence
matrix with shared item latent factors.
Hsieh \emph{et al.}~\cite{cml} propose to learn a joint metric space to encode
both user preferences and user-user and item-item similarities.
A recent work presented by Yu \emph{et al.}~\cite{walkranker} shows that
jointly modeling user-item, user-user, and item-item relations outperforms
competing methods that consider only user-item relations.
In~\cite{n2v,dw}, the higher-order proximity has been shown useful in graph
embedding methods.
In general, exploiting additional collaborative relations shows promise in
learning better representations of vertexes in an information graph.

We note that these prior arts~\cite{cofactor,cml,walkranker,n2v} share the
same core idea: using some specific methods to sample auxiliary information
from a graph to augment the data for representation learning.
However, there is a lack of a unified and efficient model that generalizes the
underlying computation and aims at recommendation problems.
For example, Liang \emph{et al.}~\cite{cofactor} consider only the item-item
similarities but no other collaborative relations; Yu \emph{et
al.}~\cite{walkranker} consider only ranking-based loss functions but not
rating-based ones.
Higher-order proximity is exploited in~\cite{dw,n2v,bine}, which however deal with
the general graph embedding problem not the recommendation one.
Moreover, the model presented by~\cite{cml} fails to manage large-scale
user-item associations~\cite{lrml}.

To address this discrepancy, in this paper we present \emph{collaborative
similarity embedding} (CSE), a unified representation learning framework for
collaborative recommender systems with the aim of modeling the direct and
in-direct edges of user-item interactions in a simple and effective way. 
CSE involves a \emph{direct similarity} embedding module for modeling
user-item associations as well as a \emph{neighborhood similarity} embedding
module for modeling user-user and item-item similarities.
The former module provides the flexibility to implement various types of modeling
techniques for user-item associations, whereas the later module models
user-user and item-item relations via $k$-order neighborhood proximity.
To simultaneously manage the two modules, we introduce \emph{triplet embedding}
into the proposed framework to ideally model user-user, item-item clustering
and user-item relations in a single and joint-learning model, while most prior
arts use only one or two embedding mappings in their methods.
Moreover, the two sub-modules are fused by a carefully designed sampling
technique for scalability and flexibility
For scalability, the space complexity and time of convergence are both only
linear with respect to the number of observed user-item associations.
In addition, with the proposed sampling techniques, CSE provides the
flexibility to shape different relation distributions in its optimization.

Extensive experiments were conducted on eight recommendation datasets that
cover different user-item interaction types, levels of data sparsity, and data
sizes.
We compare the performance of CSE with classic methods such as matrix
factorization (MF) and Bayesian personalized ranking (BPR)~\cite{bpr}, recent
methods that incorporate user-user and/or item-item
relations~\cite{cml,cofactor,walkranker}, as well as several general graph
embedding methods~\cite{dw,bine}.
The evaluation shows that CSE outperforms the competing methods for seven out
of the eight datasets. 

In summary, we propose a simple yet effective representation learning framework
aiming at making the best use of information embedded in observed user-item
associations for recommendation.
Our framework advances the state-of-the-art recommendation algorithms along the
following five dimensions.
\begin{enumerate}
    \item
      The CSE serves as a generalized framework that models comprehensive
      pairwise relations among users and items with a unified objective
      function in a simple and effective manner.
    \item 
      The proposed sampling technique enables the suitability of CSE for
      large-scale user-item recommendations.
      \item
        We provide model analyses for flexibility, scalability and time
        and space complexity of the proposed CSE.
    \item
      We report extensive experiments over eight recommendation datasets
      covering different user-item interaction types, levels of data sparsity,
      and data sizes, demonstrating the robustness, efficiency, and
      effectiveness of our framework.
    \item
      For reproducibility, we share the source code of CSE online at a GitHub
        repo,\footnote{ \url{https://github.com/cnclabs/proNet-core}
      } by which the learning process can be done within an hour for
  each dataset performed in this work.
\end{enumerate}

The rest of this paper is organized as follows.
In Section~\ref{sec:CSE}, we present the proposed CSE framework in detail,
including the problem definition, the two similarity embedding modules, and the
sampling techniques.
We then provide model analyses for flexibility, scalability and time and space
complexity of the proposed CSE in Section~\ref{sec:analysis}.
Experimental results are provided and discussed in Section~\ref{sec:exp}.
Section~\ref{sec:conclude} concludes the paper.

\section{Proposed CSE Framework}\label{sec:CSE}

\paragraph{Problem Formulation}

A recommender system provides a list of ranked items to users based on
their historical interactions with items.
Let $U$ and $I$ denote the sets of users and items, respectively.
User-item associations can be presented as a bipartite graph $G = (V,E)$, where
$V=\{v_{1},\ldots, v_{|V|}\}= U \cup I$, and $E$ represents the set of observed
user-item associations.
Note that for explicit rating data, the weights of the user-item preference
edges can be positive real numbers, whereas for implicit interactions, the
bipartite graph becomes a binary graph.
The goal of the CSE framework is to obtain an embedding matrix
$\Phi\in \mathbb{R}^{|V|\times d}$ that maps each user and item into a
$d$-dimensional embedding vector for item recommendation;
that is, with the learned embedding matrix $\Phi$, for a user $v_i\in U$, the
proposed framework generates the top-$N$ recommended items via computing the
similarity between the embedding vector of the user, i.e., $\Phi_{v_i}$, and
those of all items, i.e., $\Phi_{v_j}$ for all $v_j\in I$, where $\Phi_{v_x}$
denotes the row vector for vertex $v_x\in V$ from matrix $\Phi$. 
\paragraph{Framework Overview}

Figure~\ref{fig:overview} provides an overview of CSE.
In the figure, CSE consists of two similarity embedding modules: a direct
similarity embedding ({DSEmbed}) module to model user-item associations, and a
neighborhood similarity embedding ({NSEmbed}) module to model
user-user and item-item similarities.
The {DSEmbed} model provides the flexibility to implement two mainstream types
of modeling techniques: rating-based and ranking-based models to preserve
direct proximity of user-item associations; {NSEmbed}, in turn, models
user-user and item-item relations using the contexts within a $k$-step random
walk, as shown in Fig.~\ref{fig:overview}(b), to preserve $k$-order
neighborhood proximity between users and items.
To minimize the sum of the losses from DSEmbed and NSEmbed modules, which
are denoted as $\mathcal{L}_{DS}$ and $\mathcal{L}_{NS}$ respectively, the
objective function of the proposed framework is designed as
\begin{equation*}
\mathcal{L}= \mathcal{L}_{DS} + \lambda\mathcal{L}_{NS},
\end{equation*}
where $\lambda$ controls the balance between the two losses.
The rationale behind this design is that $\mathcal{L}_{DS}$ controls the
optimization of the embedding vectors towards preserving direct user-item
associations, and $\mathcal{L}_{NS}$ encourages users/items sharing similar
neighbors to be close to one another in the learned embedding space.

\subsection{Direct Similarity Embedding (DSEmbed) Module}

\begin{definition}{{\bf (Direct Proximity)} \rm
   Given a bipartite graph $G = ({V, E})$, the \emph{direct proximity} between
   a user $v_i\in U$ and an item $v_j\in I$ is represented by the presence of
   an edge $(v_i,v_j)\in E$ between these two vertices. If there is no edge
   between user $v_i$ and item $v_j$, then their direct proximity is defined as
 0.}
 \label{def:direct}
\end{definition}

The DSEmbed module is designed to model the direct proximity of the user-item
associations defined in Definition~\ref{def:direct}.
For a rating-based approach, the objective is to find the embedding matrix
$\Phi$ that maximizes the log-likelihood function of observed user-item pairs:
\begin{equation}
\begin{aligned}
&    \arg\max_{\Phi} \sum_{(v_i, v_j)\in E} \log p(v_i, v_j|\Phi) \\
=&   \arg\min_{\Phi} \sum_{(v_i, v_j)\in E} - \log p(v_i,v_j|\Phi).
\end{aligned}
\label{eq:log_reg}
\end{equation}

In contrast, a ranking-based approach cares more about whether we can predict
stronger association between a `positive' user-item pair $(v_i,v_j)\in E$ than
a `negative' user-item pair $(v_i,v_k)\in\bar{E}$~\cite{bpr}, where $\bar{E}$
denotes the set of edges for all the unobserved user-item associations.
This can be approached by maximizing the log-likelihood function of observed
user-item pairs over unobserved user-item pairs for each user:
\begin{equation}
\begin{aligned}
&   \arg\max_{\Phi} \sum_{ (v_i, v_j,v_k) }\log p(v_j >_i v_k|\Phi) \\
&=  \arg\min_{\Phi} \sum_{ (v_i, v_j,v_k) } - \log p(v_j >_i v_k|\Phi),
\end{aligned}
\label{eq:log_rank}
\end{equation}
where $v_i\in U$ and $v_j,v_k\in I$, and $>_i$ indicates that user $v_i$ prefers
item $v_j$ over item $v_k$.
In the above two equations, $p(v_i, v_j |\Phi)$ and $p(v_j >_i
v_k|\Phi)$ is calculated by 
\begin{equation*}
p(v_i, v_j |\Phi) = 
\sigma\left( \Phi_{v_{i}} \cdot \Phi_{v_{j}}\right),
\end{equation*}
and
\begin{equation*}
p(v_j >_i v_k|\Phi) = 
\sigma\left( \Phi_{v_{i}} \cdot \Phi_{v_{j}}
- \Phi_{v_{i}} \cdot \Phi_{v_{k}} \right),
\end{equation*}
respectively, and $\sigma(\cdot)$ denotes the sigmoid function.

\subsection{Neighborhood Similarity Embedding (NSEmbed) Module}\label{sec:nsembed}

 \begin{definition}{{\bf ($k$-Order Neighborhood Proximity)} \rm
   Given a bipartite graph $G = ({V, E})$ representing the observed user-item
   associations of the set of users and items in $V=U\cup I$, the 
   \emph{$k$-order neighborhood proximity} of a pair of users (or items) is defined
   as the similarity between their neighborhood network structures retrieved by
   $k$-step random walks.  Mathematically speaking, given the $k$-order neighborhood
   structures of a pair of users (or items), $v_i, v_j\in U$ (or $v_i, v_j\in I$,
   respectively), which are denoted as two sets of neighbor nodes $N_{v_i}$ and $N_{v_j}$,
   with $|N_{v_i}|=|N_{v_j}|=k$, the $k$-order neighborhood proximity between~$v_i$
   and $v_j$ is decided by the similarity between these two sets $N_{v_i}$ and $N_{v_j}$.
   If there are no shared neighbors between $v_i$ and $v_j$, the neighborhood
 proximity between them is 0.}
 \label{def:neighbor}
\end{definition}

The NSEmbed module is designed to model $k$-order neighborhood proximity for
capturing user-user and item-item similarities.
Given a set of neighborhood relations for users (or items) $S_U = \{(v_i,v_j)|\
  \forall v_i\in U, v_j\in N_{v_i}\}$ (or $S_I= \{(v_i,v_j)|\ \forall v_i\in I,
    v_j\in N_{v_i}\}$, respectively), the NSEmbed module seeks a set of
    embedding matrices $\Phi,\Phi^{UC},\Phi^{IC}\in \mathbb{R}^{|V|\times d}$
    that maximizes the likelihood of all pairs in $S_U$ (or $S_I$,
    respectively), where $\Phi$ is a \emph{vertex mapping matrix} akin to that
    used in the DSEmbed module, and $\Phi^{UC}$ and $\Phi^{IC}$ are two
    \emph{context mapping matrices}.
Note that each vertex (representing a user or an item) plays two roles for
modeling the neighborhood proximity:~1) the vertex itself and 2) the
context of other vertices~\cite{dw,line,n2v,app,hpe}.
With this design, the embedding vectors of vertices that share similar contexts
are thus closely located in the learned vector space.
Therefore, the maximization of the likelihood function can be defined as
\begin{eqnarray*}
\arg\max_{\Phi,\Phi^{UC},\Phi^{IC}} &
    \prod_{(v_{i}, v_{j}) \in S_U} p(v_{j}|v_{i};\Phi;\Phi^{UC}) \notag\\
&+  \prod_{(v_{i}, v_{j}) \in S_I} p(v_{j}|v_{i};\Phi;\Phi^{IC}).
\end{eqnarray*}

Similar to Eqs.~(\ref{eq:log_reg}) and (\ref{eq:log_rank}), the above
objective function becomes
\begin{eqnarray}
\arg\min_{\Phi,\Phi^{UC},\Phi^{IC}} &
    \sum_{(v_{i}, v_{j}) \in S_U} - \log p(v_{j}|v_{i};\Phi;\Phi^{UC}) \notag\\
&+  \sum_{(v_{i}, v_{j}) \in S_I} - \log p(v_{j}|v_{i};\Phi;\Phi^{IC}),
\label{eq:log_similar}
\end{eqnarray}
where
\begin{equation}
p(v_{j}|v_{i};\Theta) = 
\begin{cases}
  \sigma\left( \Phi_{v_{i}} \cdot \Phi^{UC}_{v_{j}}\right)
& \quad \text{if }\, v_{i} \in U, \\
\sigma\left( \Phi_{v_{i}} \cdot \Phi^{IC}_{v_{j}}\right)
& \quad \text{if }\, v_{i} \in I.
\end{cases}
\end{equation}

It is worth mentioning that most prior arts use only one or two embedding
mappings; while the former approach fails to consider high-order 
neighbors (e.g.,~\cite{line,libmf,mf,cdl,trustsvd}), 
the later one cannot model user-user, item-item, and user-item relations
simultaneously (e.g.,~\cite{dw,wl,hpe,n2v,app,i2v}).
Our newly designed triplet embedding solution (i.e., $\Theta=\{\Phi, \Phi^{UC},
\Phi^{UC}\}$) can ideally model user-user, item-item clustering and user-item
relations in a single and joint-learning model.


\subsection{Sampling-based Expectation Loss}
\label{sec:sample}

In order to minimize the above objective functions, we need to go through
all the pairs in $E$ for Eq.~(\ref{eq:log_reg}), $E$ and $\bar{E}$ for
Eq.~(\ref{eq:log_rank}), and $S_U$ and $S_I$ for Eq.~(\ref{eq:log_similar}),
to compute all the pairwise losses. This is not feasible in real-world
recommendation scenarios as the complexity is $\mathcal{O}(|V|\times|V|)$. 
To address this, we propose a sampling technique to work in tandem with the above
two modules to enhance CSE's scalability and flexibility in learning user and
item representations from large-scale datasets.

In CSE, the DSEmbed and NSEmbed modules are fused with the shareable
data sampling technique described below.
For each parameter update, we first sample an observed user-item 
pair $(v_i, v_j)\in E$, as shown as \texttt{U1} and \texttt{I1} in
Fig.~\ref{fig:overview}(b), where $v_i\in U$ and $v_j\in I$.
Then, we search for the $k$-order neighborhood structures of user $v_i$ and
item $v_j$ via the $k$-step random walks.
To improve computational efficiency, we use negative sampling~\cite{line}.
Consequently, for a rating-based approach (see Eq.~(\ref{eq:log_reg})),
the expected sampled loss of the DSEmbed module can be re-written as
\begin{eqnarray}
\mathcal{L}_{DS} &=&
    \mathbb{E}_{(v_i, v_j) \sim E} \left[\,-\log p(v_i,v_j|\Phi) \,\right] + \notag
    \\
&&  \sum_{M} \mathbb{E}_{(v_k, v_h) \sim \bar{E}} \left[\,\log p(v_k,v_h|\Phi) \,\right]\,,
\label{eq:DS_regr}
\end{eqnarray}
where $M$ denotes the number of negative pairs adopted.
For a ranking-based approach (see Eq.~(\ref{eq:log_rank})), the DSEmbed module
can be re-written as
\begin{equation}
\mathcal{L}_{DS} 
= \mathbb{E}_{(v_i, v_k) \sim \bar{E}} \left[ \mathbb{E}_{(v_i, v_j) \sim {E}}
    \left[\,-\log p(v_j >_i v_k|\Phi) \right] |\, v_i \,\right]\,. \\
\label{eq:DS_rank}
\end{equation}
Note that for the ranking-based approach, there is no need to explicitly
include $M$ negative sample pairs as this kind of method naturally involves negative
pairs from $\bar{E}$.
Similarly, given a user or an item vertex $v_{i}$, its $k$-order
neighborhood structure $N_{v_i}$ is composed of nodes in the $k$-step random
walks surfing on $G$, $\mathcal{W}_{v_i} = (\mathcal{W}_{v_i}^{0},\mathcal{W}_{v_i}^{1},
\mathcal{W}_{v_i}^{2}, \dots, \mathcal{W}_{v_i}^{k})$, where the vertex for
$\mathcal{W}^j_{v_i}$ is randomly chosen from the neighbors of the vertex $v$
given $\mathcal{W}_{v_i}^{j-1}=v$ and $\mathcal{W}_{v_i}^{0}=v_i$.
The expected sampled loss of the NSEmbed module can be re-written as
\begin{eqnarray}
\mathcal{L}_{NS} 
&=&
\notag\mathbb{E}_{(v_{i},v_{j}) \sim S_U}
\left[\,-\log p(v_{j}|v_{i};\Phi;\Phi^{UC}) \,\right] +\notag\\
&&  \sum_{M} \mathbb{E}_{(v_{i},v_{j}) \sim \bar{E}}
\left[\, \log p(v_{j}|v_{i};\Phi;\Phi^{UC}) \,\right] +\notag\\
&&
\mathbb{E}_{(v_{i},v_{j}) \sim S_I}
\left[\,-\log p(v_{j}|v_{i};\Phi;\Phi^{IC}) \,\right] +\notag\\
&&  \sum_{M} \mathbb{E}_{(v_{i},v_{j}) \sim \bar{E}}
\left[\, \log p(v_{j}|v_{i};\Phi;\Phi^{IC}) \,\right].
\label{eq:NS}
\end{eqnarray}

Since $\mathcal{L}_{DS}$ and $\mathcal{L}_{NS}$ are described in a sampling-based
expectation form, CSE provides the flexibility for accommodating arbitrary
distributions of positive and negative data. In the following experiments, we produce
the positive data according to primitive edges distribution of given user-item graph.
As to negative sampling, we propose to directly sample the negative data from whole
data collection instead of unobserved data collection.

\subsection{Optimization}

In the optimization stage, we use asynchronous stochastic gradient
descent (ASGD)~\cite{ASGD} to efficiently update the parameters in parallel.
The model parameters are composed of the three embedding matrices $\Theta =
\{\Phi,\Phi^{UC},\Phi^{IC}\}$, each having the size $\mathcal{O}(|V|d)$.
They are updated with learning rate $\alpha$ according to 
\begin{equation}
\Theta \leftarrow \Theta - \alpha \left(
\frac{\partial\mathcal{L}_{DS}}{\partial \Theta} + \lambda \left(\frac{\partial\mathcal{L}_{NS}}{\partial \Theta} \right)
- \lambda_{V}\Vert\Phi\Vert
\right)\,, 
\label{eq:loss}
\end{equation}
where $\lambda_{V}$ is a hyper-parameter for reducing the risk of overfitting.

{
\begin{table*}%
\footnotesize
\centering
\begin{threeparttable}[t]
\resizebox{0.9\textwidth}{!}
{
\begin{tabular}{lrrrrrll}
\toprule
    Dataset & \#Users &  \#Items & \#Edges & $\frac{\#\text{Edges}}{\#\text{Users}}$ & Density & Edge type & Network type \\
\midrule
    Frappe\tnote{a}		    & 957		& 4,028		& 96,202	    & 100.52    & 2.50\%	& click count  & app-clicks		\\
    CiteULike\tnote{b}      & 5,551		& 16,980	& 210,504	    & 37.92     & 0.22\%	& like/dislike & references		\\
    Netflix\tnote{c}        & 65,533	& 17,759	& 25,120,129    & 383.32    & 2.15\%	& 5-star       & movie-ratings		\\
    MovieLens-Latest\tnote{d}    & 259,137	& 40,110	& 24,404,096    & 94.17     & 0.23\% 	& 5-star       & movie-ratings 	\\
    Last.fm-360K\tnote{e}        & 359,347	& 294,015	& 17,559,530    & 48.86     & 0.02\%	& play count   & artist-plays		\\
    Amazon-Book\tnote{f}    & 603,668	& 367,982	& 8,898,041	    & 14.73     & 0.004\%	& 5-star       & book-ratings		\\
    Epinions-Extend\tnote{g}& 755,760	& 120,492	& 13,668,319    & 18.08     & 0.02\%	& 5-star       & product-reviews	\\
    Echonest\tnote{h}       & 1,019,318 & 384,546	& 48,373,586    & 47.45     & 0.01\%	& play count   & song-plays		\\
\bottomrule
\end{tabular}
}
\caption{Statistics of the datasets considered in our experiments.}
\label{tb:data}
\begin{tablenotes}
\setlength{\columnsep}{0.8cm}
\setlength{\multicolsep}{0cm}
\begin{multicols}{2}
    \item[a] \url{http://baltrunas.info/research-menu/frappe}
    \item[b] \url{http://www.wanghao.in/CDL.htm}
    \item[c] \url{http://academictorrents.com/}
    \item[d] \url{https://grouplens.org/}
    \item[e] \url{http://www.dtic.upf.edu//~ocelma/MusicRecommendationDataset/}
    \item[f] \url{http://jmcauley.ucsd.edu/data/amazon/}
    \item[g] \url{http://www.trustlet.org/epinions.html}
    \item[h] \url{https://labrosa.ee.columbia.edu/millionsong/tasteprofile}
\end{multicols}
\end{tablenotes}
\end{threeparttable}
\end{table*}%
}

\section{Model Analysis}\label{sec:analysis}

The CSE framework not only modularizes the modeling of pairwise user-item,
user-user, and item-item relations, but also integrates them into a single
objective function through shared embedding vectors.
Together with the DSEmbed and NSEmbed sub-modules for modeling
these relations, CSE involves a novel sampling technique to 
improve scalability and flexibility.
To give a clear view of CSE, we further provide the comparison to general graph
embedding and deep learning models. 

\subsection{Scalability}
Typical factorization methods usually work on a sparse user-item matrix and do
not explicitly model high-order connections from the corresponding user-item
bipartite graph $G=(V,E)$.
Several methods, including our CSE, propose to explicitly incorporate
high-order connections for modeling user-user and item-item relations into the
recommendation models to improve performance.
As discussed in~\cite{embedmf,cofactor,w2v}, such modeling can then be seen as
conducting matrix factorization on a $|V|\times|V|$ point-wise mutual
information (PMI) matrix:
\begin{equation*}
  \text{PMI}\,(v_{i},v_{j}) = \log \left(\frac{p(v_{i},
  v_{j})}{p(v_{i})p(v_{j})}\right) - \log M \,.
\end{equation*}
Recall that $V=U\cup I$, and $M$ denotes the number of negative pairs adopted
in Eqs.~(\ref{eq:DS_regr}) and (\ref{eq:NS}).
However, given the high-order connections for modeling
user-user and item-item relations, most $\text{PMI}(v_i,v_j)$ for $v_i,v_j\in U$
(or $v_i,v_j\in I$) are nonzero and thus the PMI matrix is considered
non-sparse. Conducting matrix factorization on such a matrix is
computationally expensive in both time and space, and is therefore
infeasible in many large-scale recommendation scenarios.

To explicitly consider all the collaborative relations into our model while keeping
it practical for large-scale datasets, the CSE uses the
sampling technique together with $k$-step random walks surfing on $G$ to
preserve direct proximity of user-item associations as well as harvest the
high-order neighborhood structures of users and items.
By doing so, we approximate factorization of the corresponding PMI matrix and
thus reduce the complexity in space and the training time.

\subsection{Flexibility}\label{sec:flex}
In the CSE, the DSEmbed and NSEmbed modules are united
with the sampling technique.
In addition to improved scalability, such a sampling perspective facilitates
the shaping of different relation distributions for optimization via
different weighting schemes or sampling strategies.
Here we resort to use the perspective of KL divergence to explain this model
characteristic.
Specifically, minimizing the losses in our framework (see Eqs.~(\ref{eq:DS_regr}), 
(\ref{eq:DS_rank}), and (\ref{eq:NS})) can be related to
minimizing the KL divergence of two probability distributions~\cite{line}.
Suppose there are two distributions over the space $V\times V$:
$\hat{p}(\cdot,\cdot)$ and $p(\cdot,\cdot)$, denoting the empirical
distribution and the target distribution, respectively, we have
\begin{eqnarray}
&&\arg\min_{p} \text{KL}(\hat{p}(\cdot,\cdot),p(\cdot,\cdot)) \notag\\
&=& \arg\min_{p} - \sum_{(v_i,v_j)\in E}\hat{p}(v_i,v_j)
    \log \left(\frac{p(v_i,v_j)}{\hat{p}(v_i,v_j)}\right) \notag\\
&\propto& \arg\min_{p} -\sum_{(v_i,v_j)\in E} \hat{p}(v_i,v_j)\log p(v_i,v_j).
\label{eq:KL}
\end{eqnarray}
In Eq.~(\ref{eq:KL}), the empirical distribution $\hat{p}(\cdot,\cdot)$ can be
treated as the probability density (mass) function of the distribution in our
loss functions, from which each pair of vertices $(v_i, v_j)$ is sampled.
This indicates that applying different weighting schemes or different sampling
strategies (i.e., different $\hat{p}(\cdot,\cdot)$) in CSE shapes
different relation distributions for learning representations.


\subsection{Complexity}
The time and space complexity of the proposed method depends on the
implementation.
The training procedure of CSE framework involves a sampling step and an
optimization step.
Since all the required training pairs, including observed associations and
unobserved associations, can be derived from the user-item bipartite graph $G$,
we adopt the compressed sparse alias rows (CSAR) data structure to perform
weighted edge sampling for direct similarity and weighted random walk for
neighborhood similarity~\cite{csar}.
With CSAR data structure, sampling an edge requires only $\mathcal{O}(1)$ and
the overall demand space complexity is linearly increased with the number of
positive edges $\mathcal{O}(|E|)$.\footnote{Note that the space for the learned
  embedding matrices $\Theta=\{\Phi,\Phi^{UC},\Phi^{IC}\}$ is
  $\mathcal{O}(|V|)$ and $|V|\ll|E|$.}
As to the optimization,
SGD-based update has a closed form so that updating the embedding of a vertex
in a batch depends only on the dimension size $O(d)$.
As for the time of convergence, many studies on graph embedding as well as our
method empirically show that the required total training time for the
convergence of embedding learning is also linear in $|E|$~\cite{rare}.



\subsection{Comparison to General Graph Embedding Models}
\label{sec:ge}
General graph embedding algorithms, such as DeepWalk~\cite{dw} and
node2vec~\cite{n2v}, can be used for the task of recommendation.
Yet, we do not focus on comparing the proposed CSE with general graph embedding
models and only provide the results of DeepWalk because many prior works on
recommendation \cite{e2r,nerm} have shown that many of the baseline
methods considered in our paper outperform these general graph
embedding algorithms.
The main reason for this phenomenon is that most graph embedding methods
cluster vertices that have similar neighbors together, and thus make the users
apart from the items because user-item interactions typically form a bipartite
graph. 

\subsection{Comparison to Deep Learning Models}

Our method, and the existing methods we discussed and compared in the
experiments, focus on improving the modeling quality of user and item
embeddings that can be directly and later used for user-item recommendation
with similarity computation.
Many approximation techniques, such as approximate nearest neighbor\footnote{https://github.com/erikbern/ann-benchmarks}{(ANN)}, can
be applied to speed up the similarity computation between user and item
embeddings, which facilitates real-time online predictions and makes the
recommendation scalable to large-scale real-world datasets.
In contrast, many deep learning methods, including NCF~\cite{ncf}, DeepFM~\cite{deepfm},
etc, do not learn the directly comparable embeddings of users or items. There are a few
deep learning methods (e.g., Collaborative Deep Embedding~\cite{cde}
and DropoutNet~\cite{dropoutnet})
that can produce user and item embeddings, but to our knowledge, efficiency is still
a major concern of these methods. Therefore, improving the embedding quality is
still a critical research issue for building up a recommender system especially
when the computation power is limited in real-world application scenarios.
For readability and to maintain the focus of this work, we opt for not
comparing the deep learning methods in our paper.
It is also worth mentioning that our solution can obtain user and item
embeddings within only an hour for every large dataset listed in the paper; the
efficiency and scalability is thus one of the highlights of the proposed
method.

\section{Experiment}\label{sec:exp}

\subsection{Settings}

\subsubsection{Datasets and Preprocessing}
To examine the capability and scalability of the proposed CSE framework, we
conducted experiments on eight publicly available real-world datasets that vary
in terms of domain, size, and density, as shown in Table~\ref{tb:data}.
For each of the datasets, we discarded the users who have less than ten
associated interactions with items.
In addition, we converted each data into implicit feedback:\footnote{Note that
  in real-world scenarios, most feedback is not explicit but
  implicit~\cite{bpr}; we here converted the datasets into implicit feedback as
  most of the recent developed methods focus on dealing with such type of data.
  However, our method is not limited to binary preference since the presented
  sampling technique has the flexibility to manage arbitrary weighted edge
  distributions and rating estimation is also allowed in the
  proposed RATE-CSE.}
  1) for 5-star rating datasets, we transformed ratings higher than or equal to
  3.5 to 1 and the rest to 0; 2) for count-based datasets, we transformed
  counts higher than or equal to 3 to 1 and the rest to 0; 3) for the CiteULike
  dataset, no transformation was conducted as it is already a binary preference
  dataset.

\subsubsection{Baseline Algorithms}
We compare the performance of our model with the following eight baseline
methods: 1) POP, a naive popularity model that ranks the items by their
degrees, 2) DeepWalk~\cite{dw}, a classic algorithm of network embedding, 3)
WALS~\cite{wals}, a weighted rating-based factorization model, 4) ranking-based
factorization models: BPR~\cite{bpr}, WARP~\cite{warp}, and $K$-OS~\cite{kos},
5) BiNE~\cite{bine}, a network embedding model specialized for bipartite
networks, and 6) recent advanced models considering user-user/item-item
relations: coFactor~\cite{cofactor}, CML~\cite{cml} and
WalkRanker~\cite{walkranker}, Note that except for POP, the embedding vectors
for users and items learned by these competitors as well as by our method can
be directly used for item recommendations.
Additionally, while CML adopts Euclidean distance as the scoring
function, all other methods including ours utilize the dot product to calculate
the score of a pair of user-item embedding vectors.
The experiments for WALS and BPR were conducted using the matrix factorization
library QMF,\footnote{\url{https://github.com/quora/qmf}} and those for WARP
and $K$-OS were conducted using LightFM;\footnote{\url{https://github.com/lyst/lightfm}}
for coFactor, CML, and WalkRanker, we used the code provided by the respective authors.

\subsubsection{Experimental Setup}
For all the experiments, the dimension of embedding vectors was fixed to 100;
the values of the hyper-parameters for the compared method were decided via
implementing a grid search over different settings, and the combination that
leads to the best performance was picked.
The ranges of hyper-parameters we searched for the compared methods are listed
as follows.
\begin{itemize}
    \item learning rate: [0.0025, 0.01, 0.025, 0.1]
    \item   regularization: [0.00025, 0.001, 0.0025, 0.01, 0.025, 0.1]
    \item   training epoch: [10, 20, 40, 80, 160]
    \item   sampling time: [$20\times|E|$, $40\times|E|$, $60\times|E|$, $80\times|E|$, $100\times|E|$]
    \item   walk time: [10, 40, 80]
    \item   walk length: [40, 60, 80]
    \item   window size: [2, 3, 4, 5, 6, 8, 10]
    \item   stopping probability for random walk: [0.15, 0.25, 0.5, 0.75]
    \item   $k$-order: [1, 2, 3]
    \item   rank margin: 1 (commonly used default value)
    \item   number of negative samples: 5 (commonly used default value)
\end{itemize}

For our model, the learning rate $\alpha$ was set to 0.1, $\lambda_V$ was set to
0.025; the hyper-parameter $\lambda$ was set to 0.05 and 0.1 for
rating-based CSE and ranking-based CSE, respectively, and $k$ was set to~2
as the default value. The sensitivity of CSE parameters are additionally reported.
For each dataset, the sample time for convergence depends on the number of
non-zero user-item interaction edges and is set to $80\times|E|$.
Sensitivity analysis for $k$ and $\lambda$ and convergence analysis are
later provided in the section for convergence analyses.

{
\begin{table*}%
\footnotesize
\centering
\resizebox{\textwidth}{!}{
\begin{tabular}{lrrrrrrrrr}
\toprule
    & \multicolumn{2}{c}{Frappe} & \multicolumn{2}{c}{CiteULike} & \multicolumn{2}{c}{Netflix} & \multicolumn{2}{c}{MovieLens-Latest} \\
\midrule
                                & Recall@10 & mAP@10	& Recall@10	& mAP@10	& Recall@10	& mAP@10    & Recall@10 & mAP@10 \\
\midrule
    Pop		                    & 0.1750    & 0.0708    & 0.0270    & 0.0114    & 0.0861    & 0.0359    & 0.0882    & 0.0289        \\
    DeepWalk \cite{dw}		    & 0.0430    & 0.0256    & 0.0875    & 0.0458    & 0.0235    & 0.0112    & 0.0207    & 0.0061        \\
    WALS \cite{wals}            & 0.1632    & 0.1117    & 0.1851    & 0.0915    & 0.1214    & 0.0471    & 0.2350    & $\dagger$0.1682      \\
    BPR \cite{bpr}              & 0.2785    & 0.1550    & 0.0861    & 0.0426    & 0.1496    & 0.0757    & 0.2163    & 0.1130       \\
    WARP \cite{warp}            & 0.3012    & 0.1796    & 0.1468    & 0.0813    & $\dagger$0.1887   & $\dagger$0.1004   & $\dagger$0.2712   & 0.1651       \\
    $K$-OS \cite{kos}           & 0.3018    & 0.1914    & 0.1356    & 0.0756    & 0.1783    & 0.0868    & 0.2522    & 0.1641       \\
    BiNE \cite{bine}		    & 0.2159    & 0.1201    & 0.0422    & 0.0201    & -    & -    & -    & -        \\
    coFactor \cite{cofactor}    & 0.2110    & 0.1309    & 0.1323    & 0.0721    & -     &     -     &     -     &     -        \\
    CML \cite{cml}              & $\dagger$0.3311   & 0.1958    & 0.1740    & 0.1008    & 0.1035	& 0.0444    & 0.1109    & 0.0957        \\ 
    WalkRanker \cite{walkranker}& 0.3286    & $\dagger${\bf 0.2099}   & $\dagger$0.2059   & $\dagger$0.1192   & 0.1090	& 0.0483    & 0.1307    & 0.0351        \\
\midrule
    RATE-CSE	                & {\bf 0.3347}    & 0.2047    & *{\bf 0.2362} & *{\bf 0.1452}   & *0.2014	& *0.1039   & *{\bf 0.3225}   & *{\bf 0.1990}      \\
    Improv. (\%)                & $+$1.0\%  & $-$2.4\%  & $+$14.7\%   & $+$21.9\%   & $+$6.7\%    & $+$3.5\%    & $+$18.9\%   & $+$18.3\%        \\
\midrule
    RANK-CSE	                & 0.3155    & 0.2005    & 0.1993    & *0.1228   & *{\bf 0.2156}   & *{\bf 0.1202}   & *0.3094   & *0.1902      \\
    Improv. (\%)                & $-$4.7\%  & $-$4.4\%  & $-$3.2\%  & $+$3.0\%  & $+$14.2\% & $+$19.7\% & $+$14.1\% & $+$13.1\%      \\
\end{tabular}
}

\resizebox{\textwidth}{!}{
\begin{tabular}{lrrrrrrrrr}
\toprule
    & \multicolumn{2}{c}{Last.fm-360K} & \multicolumn{2}{c}{Amazon-Book} & \multicolumn{2}{c}{Epinions-Extend} & \multicolumn{2}{c}{Echonest}\\
\midrule
                                & Recall@10	& mAP@10	& Recall@10	& mAP@10	& Recall@10	& mAP@10    & Recall@10 & mAP@10 \\
\midrule
    Pop                         & 0.0309    & 0.0133    & 0.0053    & 0.0015    & 0.0450    & 0.0246    & 0.0257    & 0.0104\\
    WALS \cite{wals}            & 0.1621    & 0.0857    & $\dagger$0.0540   & $\dagger$0.0227   & 0.1479    & 0.0634    & 0.1287    & $\dagger$0.0638\\
    BPR \cite{bpr}              & 0.1120    & 0.0545    & 0.0248    & 0.0119    & 0.1126    & 0.0579    & 0.0499    & 0.0210\\
    WARP \cite{warp}            & 0.1556    & 0.0832    & 0.0457    & 0.0199    & $\dagger$0.1509   & $\dagger$0.0775   & 0.1001    & 0.0447\\
    $K$-OS \cite{kos}           & $\dagger$0.1641   & $\dagger$0.0888   & 0.0511    & 0.0215    & 0.1493    & 0.0766    & $\dagger$0.1249   & 0.0597\\
    CML \cite{cml}              & 0.0496    & 0.0199    & 0.0129    & 0.0052    & 0.1171    & 0.0629    & 0.0357    & 0.0195\\
    WalkRanker \cite{walkranker}& 0.0233    & 0.0088    & 0.0080    & 0.0036    & 0.0560    & 0.0289    & 0.0309    & 0.0133\\
\midrule
    RATE-CSE                    & *0.1687   & *0.0909   & 0.0540    & 0.0240    & *0.1659   & 0.0788    & 0.1260    & 0.0605\\
    Improv. (\%)                &  $+$2.8\% & $+$2.3\%  & $+$0.0\%  & $+$5.7\%  & $+$9.9\%  & $+$1.7\%  & $+$0.8\%  & $-$0.5\%\\
\midrule
    RANK-CSE                    & *{\bf 0.1762}   & *{\bf 0.0970}   & *{\bf 0.0625}   & *{\bf 0.0274}   & *{\bf 0.1767}   & *{\bf 0.0921} & *{\bf 0.1358}   & *{\bf 0.0679}   \\
    Improv. (\%)                &  $+$8.2\% & $+$14.4\% & $+$15.7\% & $+$20.7\% & $+$17.0\% & $+$20.2\% & $+$8.7\%  & $+$6.4\% \\
\bottomrule
\end{tabular}
}
\caption{Recommendation performance. The $\dagger$ symbol indicates the best
  performing method among all the baseline methods; `*' and `\%Improv.' denote
  statistical significance at $p<0.01$ with a paired $t$-test and the
  percentage improvement of the proposed method, respectively, with respect to
  the best performing baseline.}
\label{tb:rec}
\end{table*}%
}

\begin{figure*}
\centering
\begin{subfigure}[b]{0.255\textwidth}
    \includegraphics[width=\textwidth,height=0.75\textwidth]{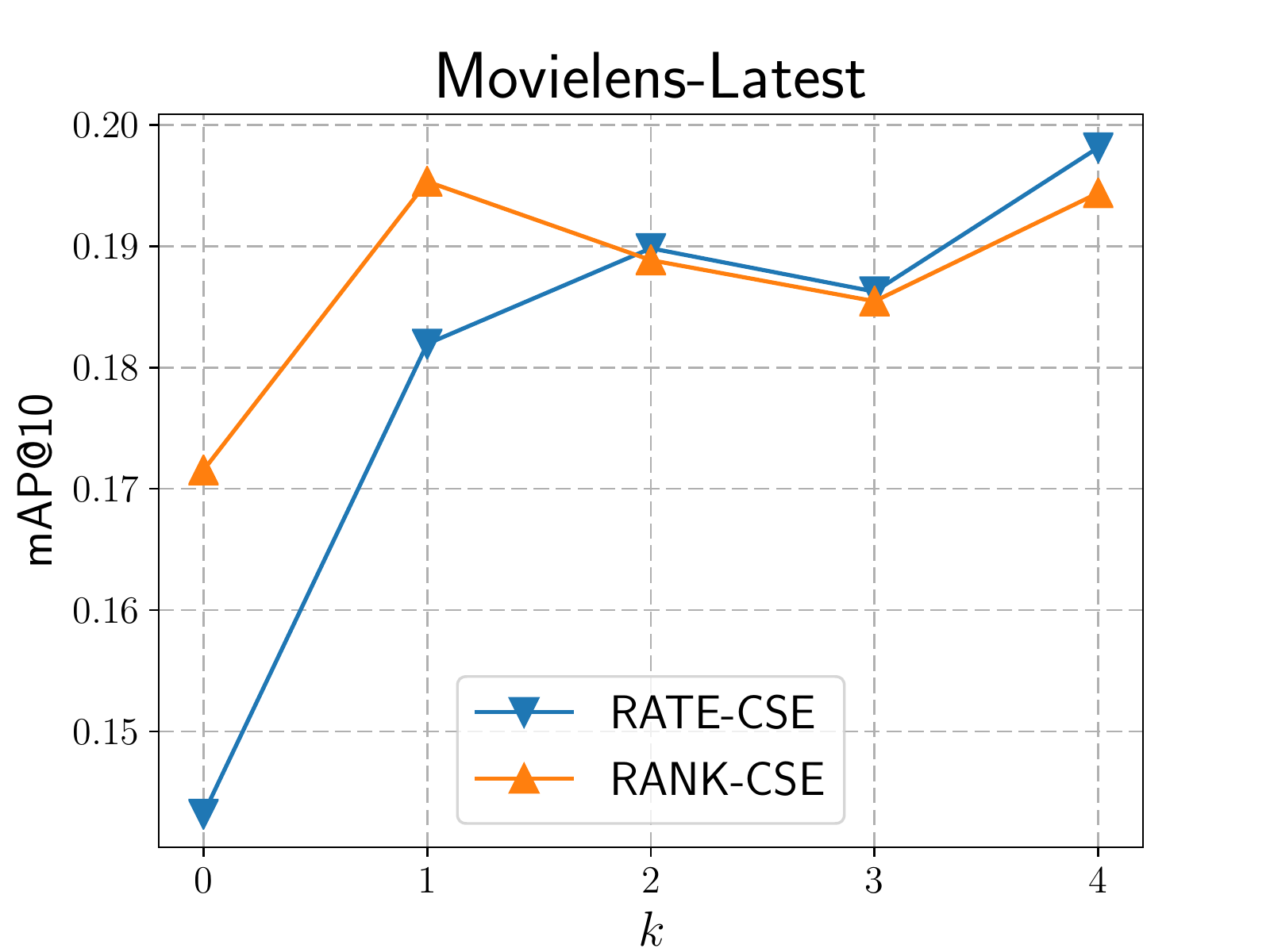}
\end{subfigure}
\hskip -2.5ex
\begin{subfigure}[b]{0.255\textwidth}
    \includegraphics[width=\textwidth,height=0.75\textwidth]{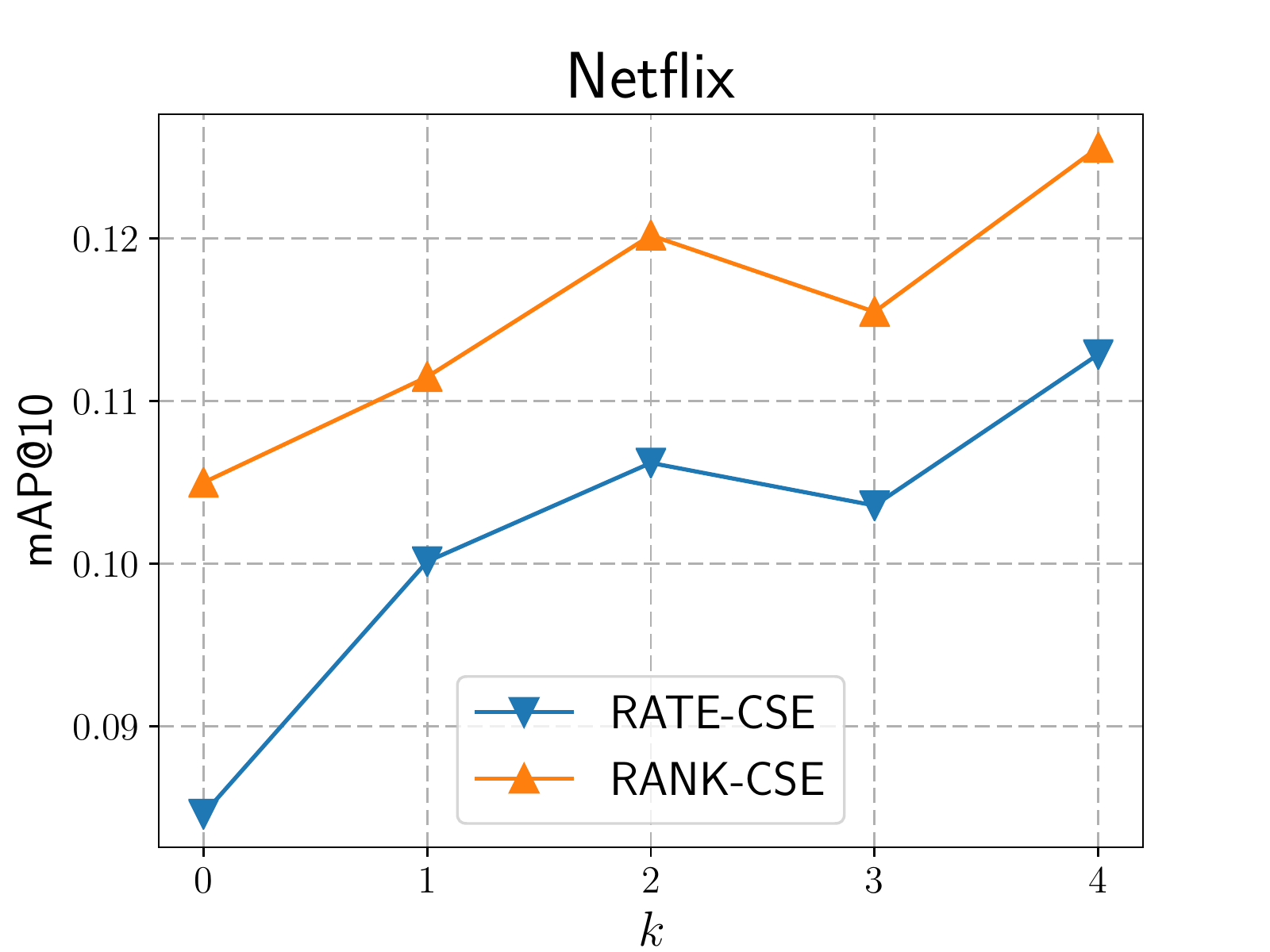}
\end{subfigure}
\hskip -2.5ex
\begin{subfigure}[b]{0.255\textwidth}
    \includegraphics[width=\textwidth,height=0.75\textwidth]{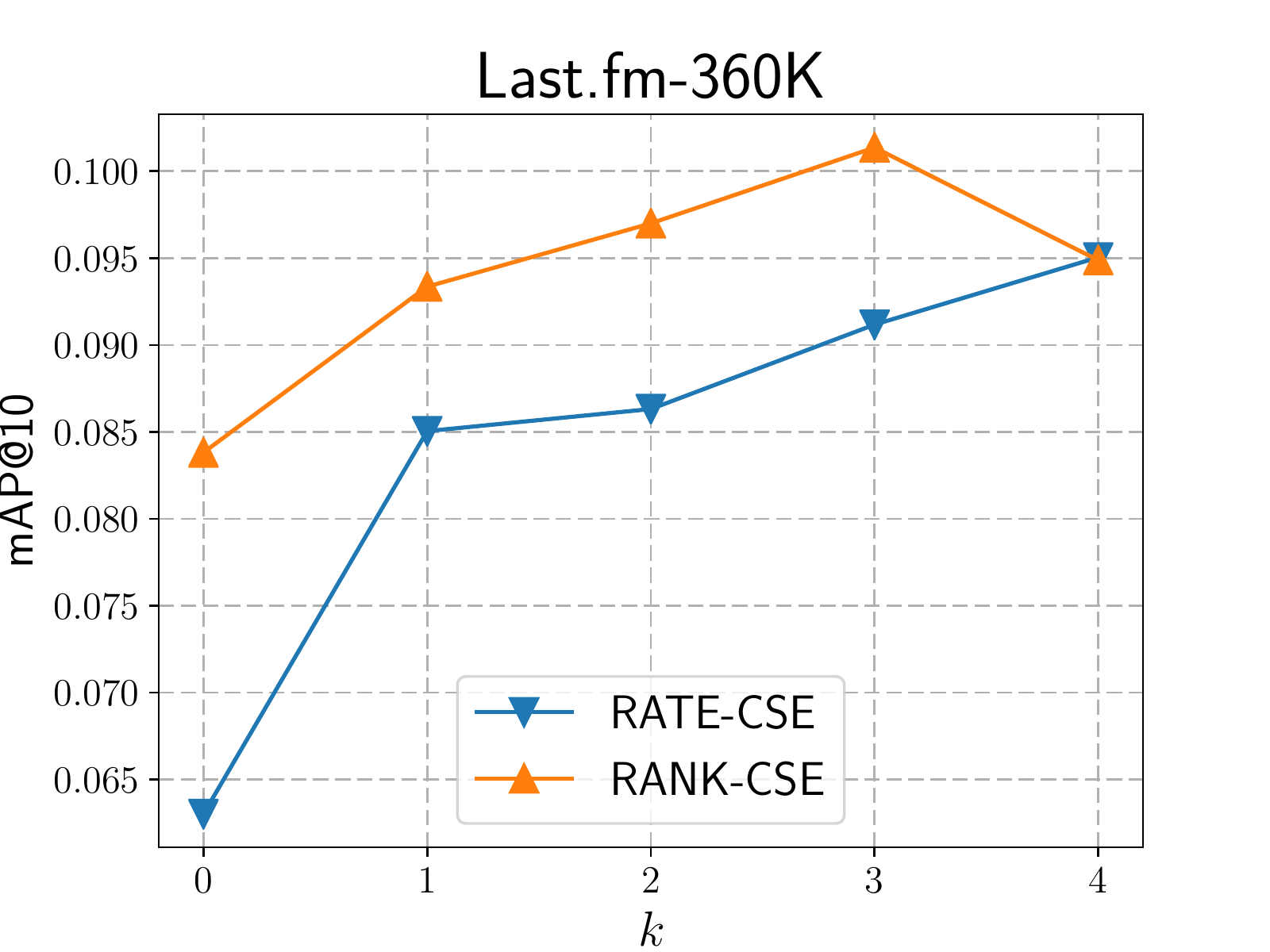}
\end{subfigure}
\hskip -2.5ex
\begin{subfigure}[b]{0.255\textwidth}
    \includegraphics[width=\textwidth,height=0.75\textwidth]{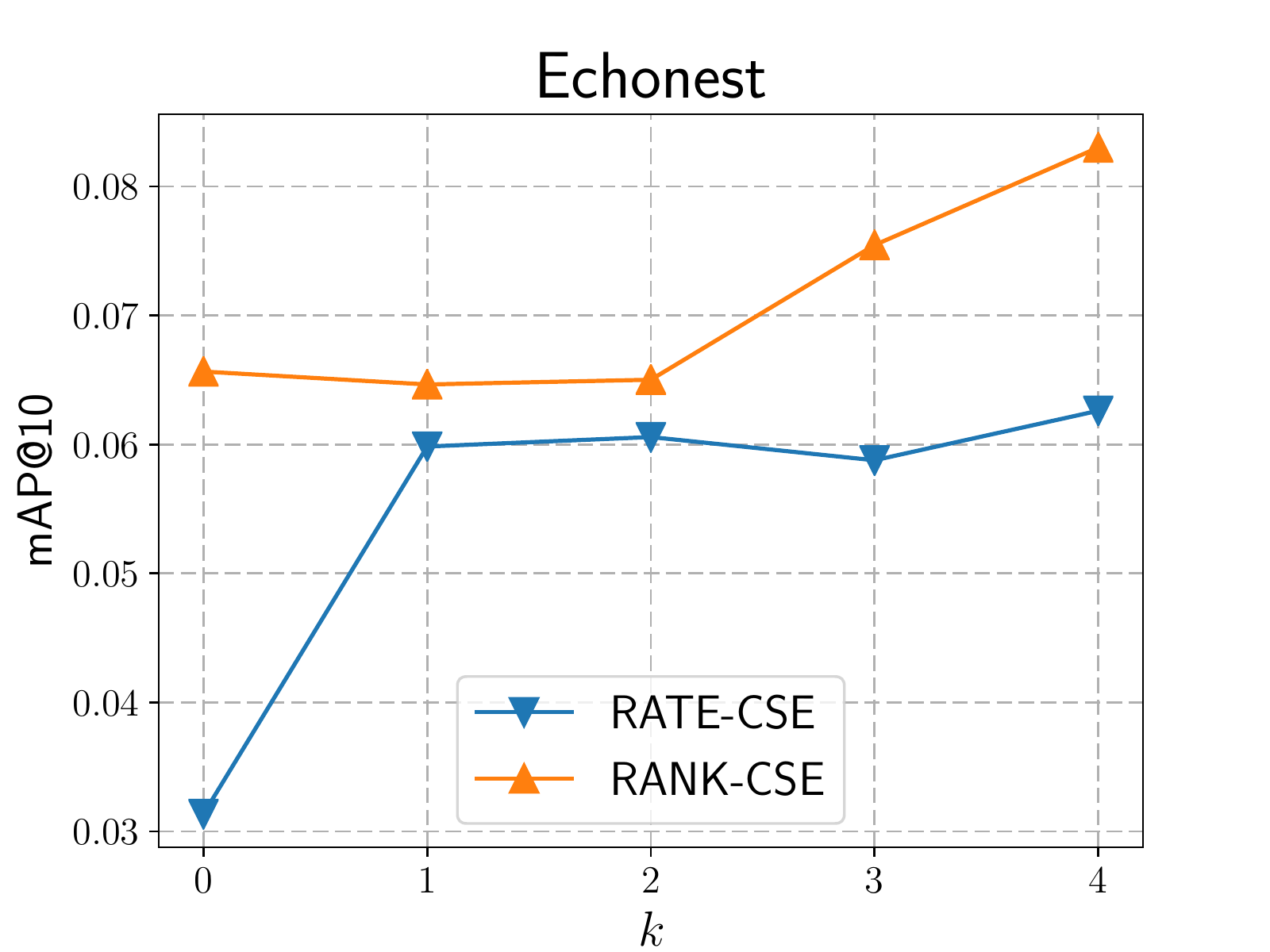}
\end{subfigure}

\begin{subfigure}[b]{0.255\textwidth}
    \includegraphics[width=\textwidth,height=0.75\textwidth]{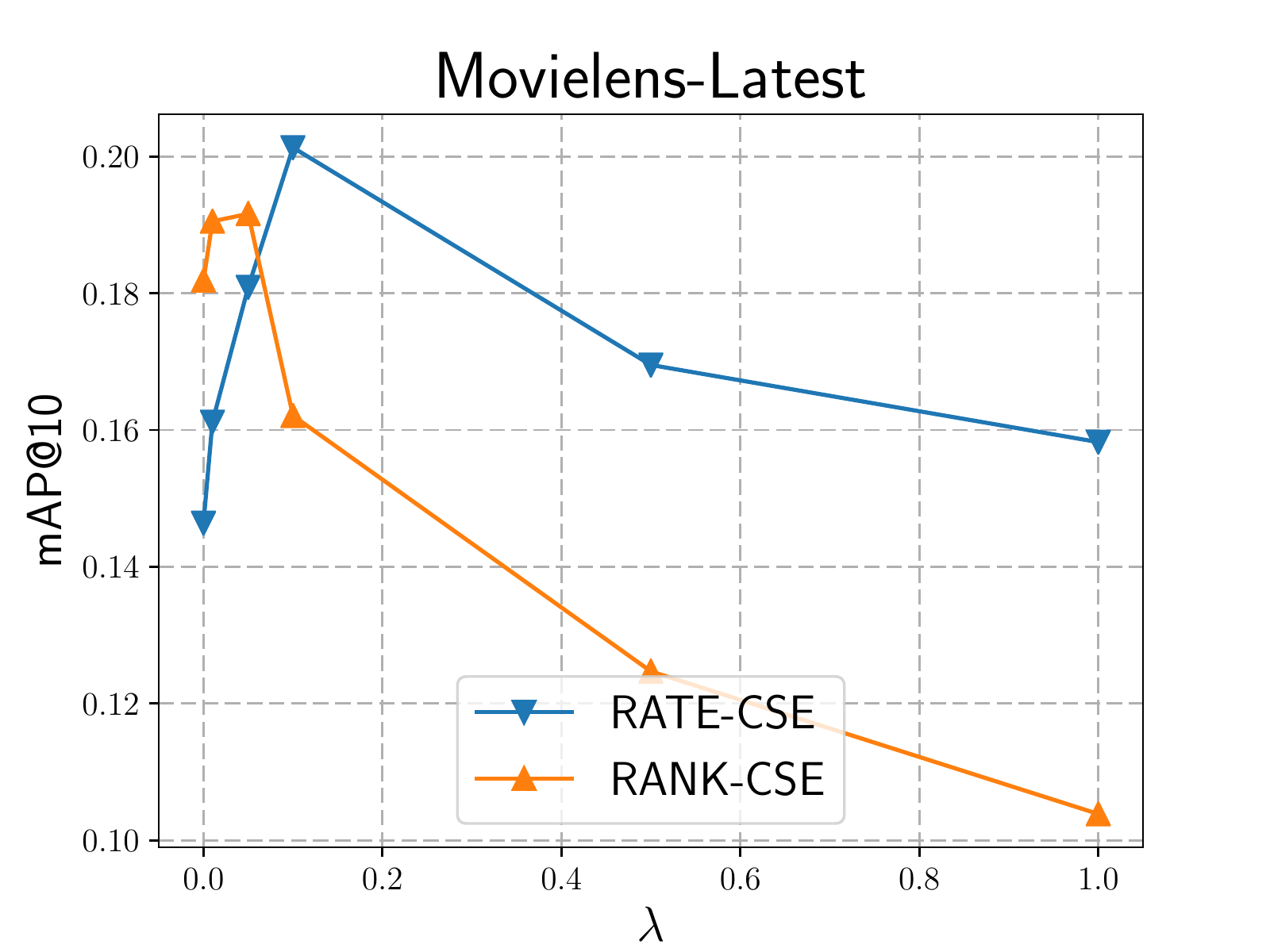}
\end{subfigure}
\hskip -2.5ex
\begin{subfigure}[b]{0.255\textwidth}
    \includegraphics[width=\textwidth,height=0.75\textwidth]{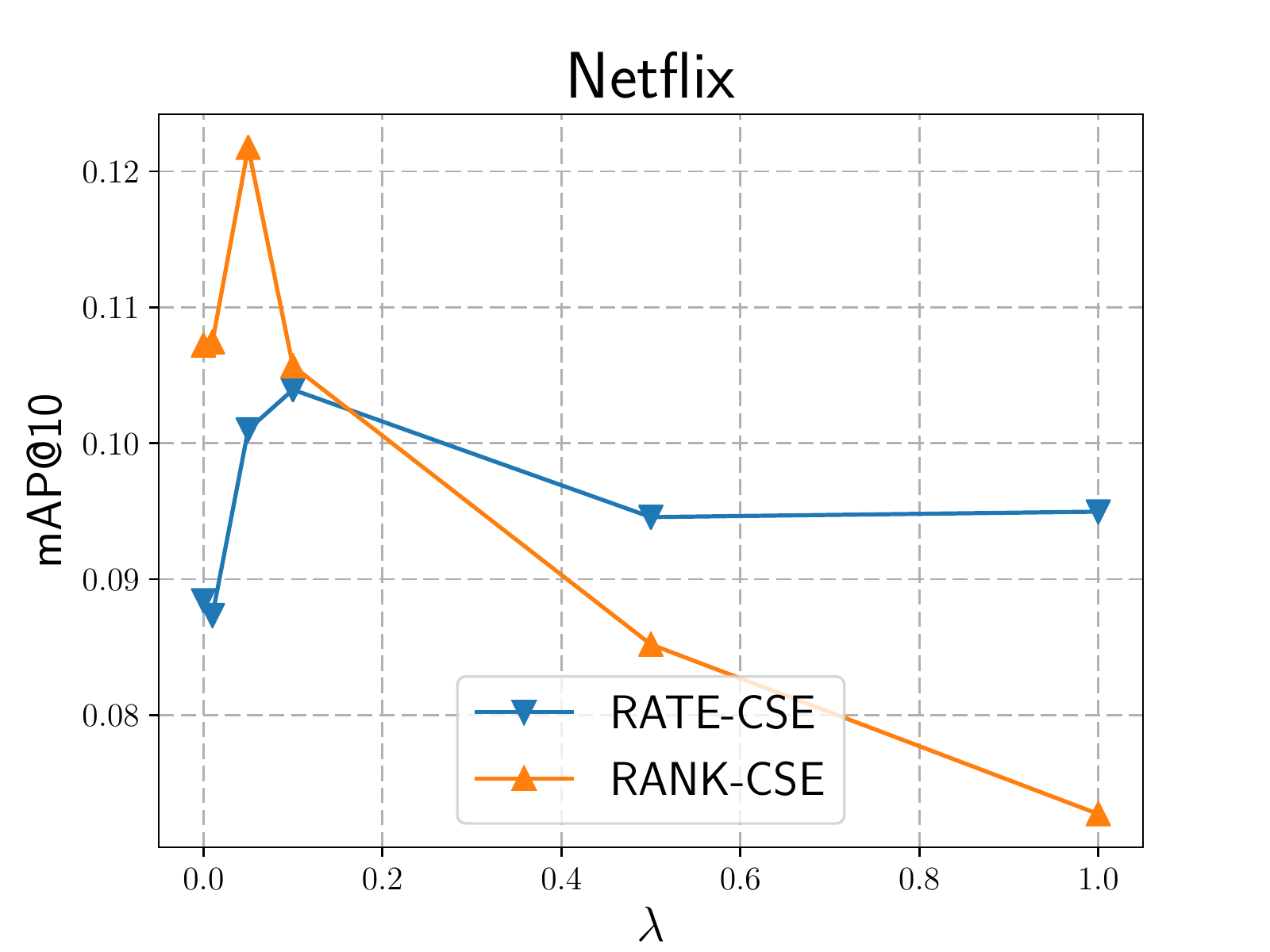}
\end{subfigure}
\hskip -2.5ex
\begin{subfigure}[b]{0.255\textwidth}
    \includegraphics[width=\textwidth,height=0.75\textwidth]{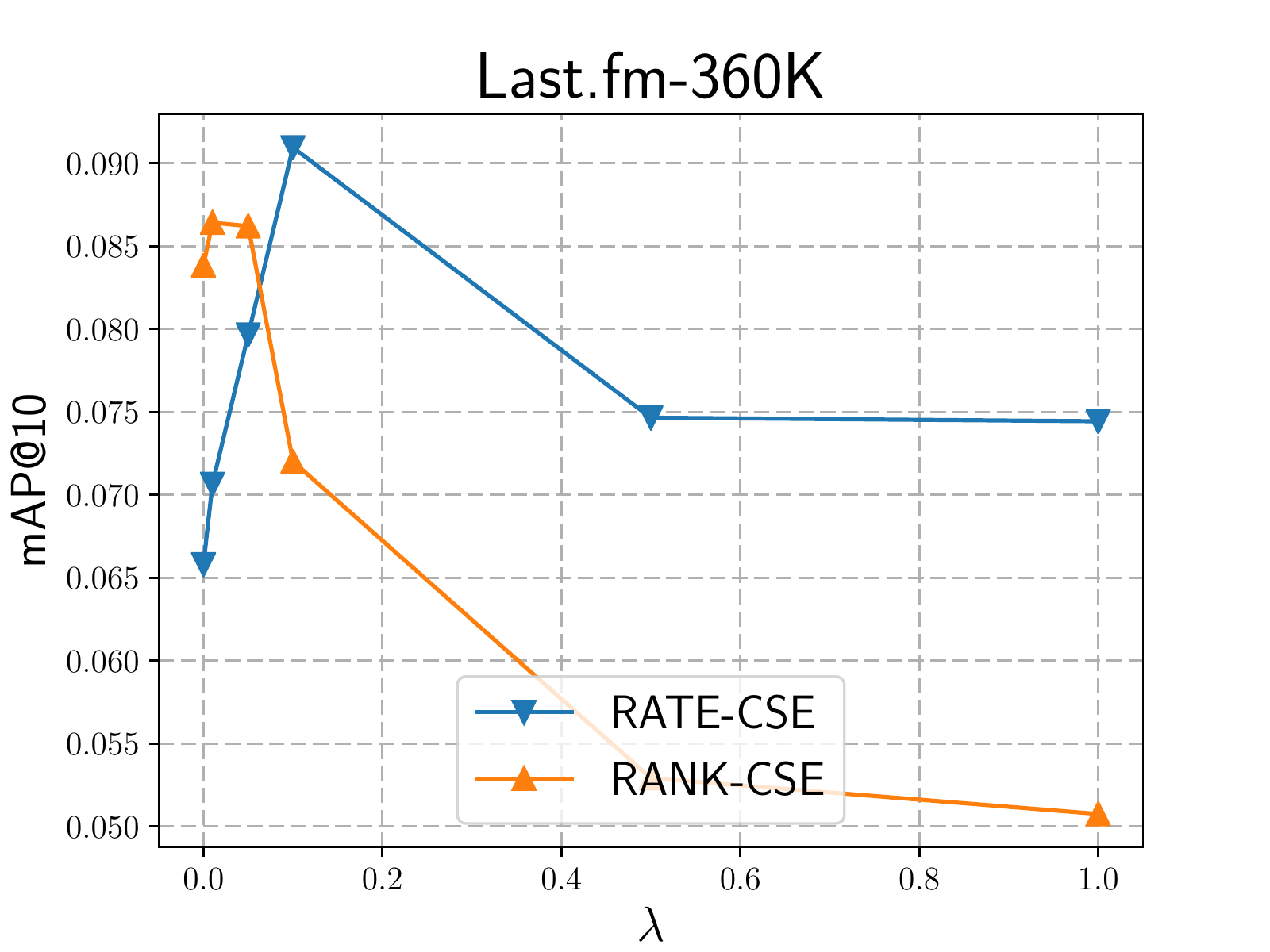}
\end{subfigure}
 \hskip -2.5ex
\begin{subfigure}[b]{0.255\textwidth}
    \includegraphics[width=\textwidth,height=0.75\textwidth]{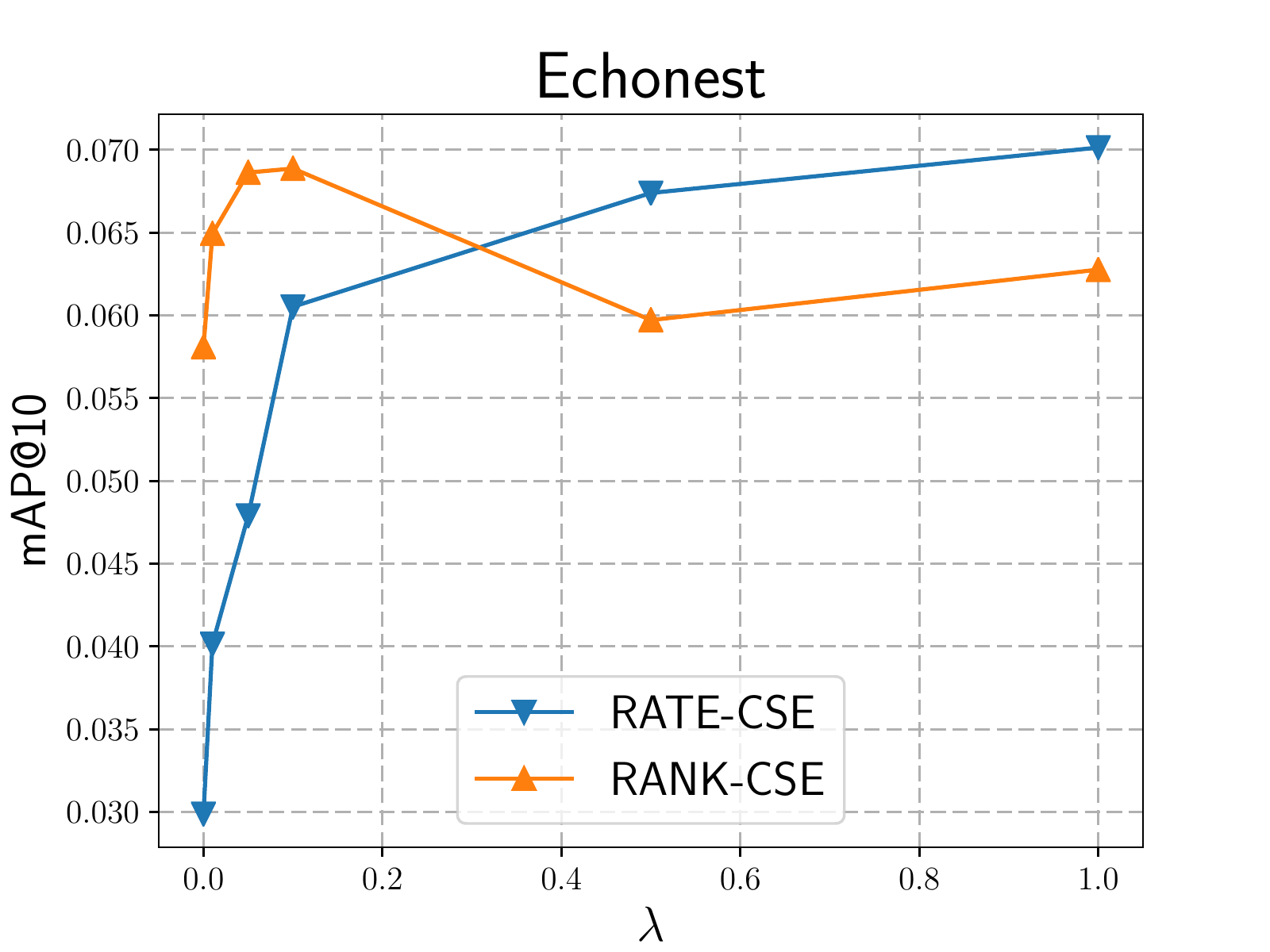}
\end{subfigure}
  
\begin{subfigure}[b]{0.255\textwidth}
    \includegraphics[width=\textwidth,height=0.75\textwidth]{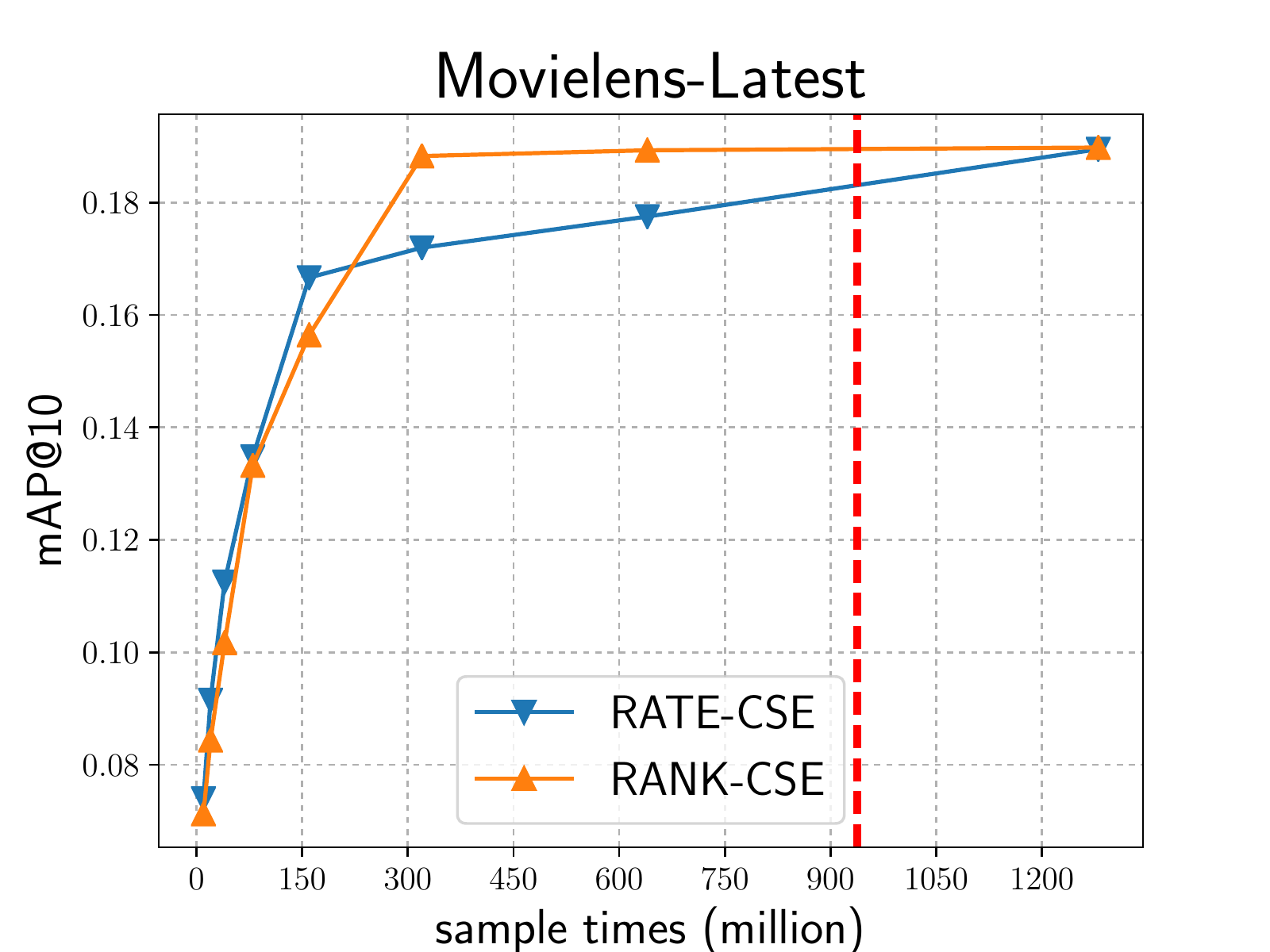}
\end{subfigure}
\hskip -2.5ex
\begin{subfigure}[b]{0.255\textwidth}
    \includegraphics[width=\textwidth,height=0.75\textwidth]{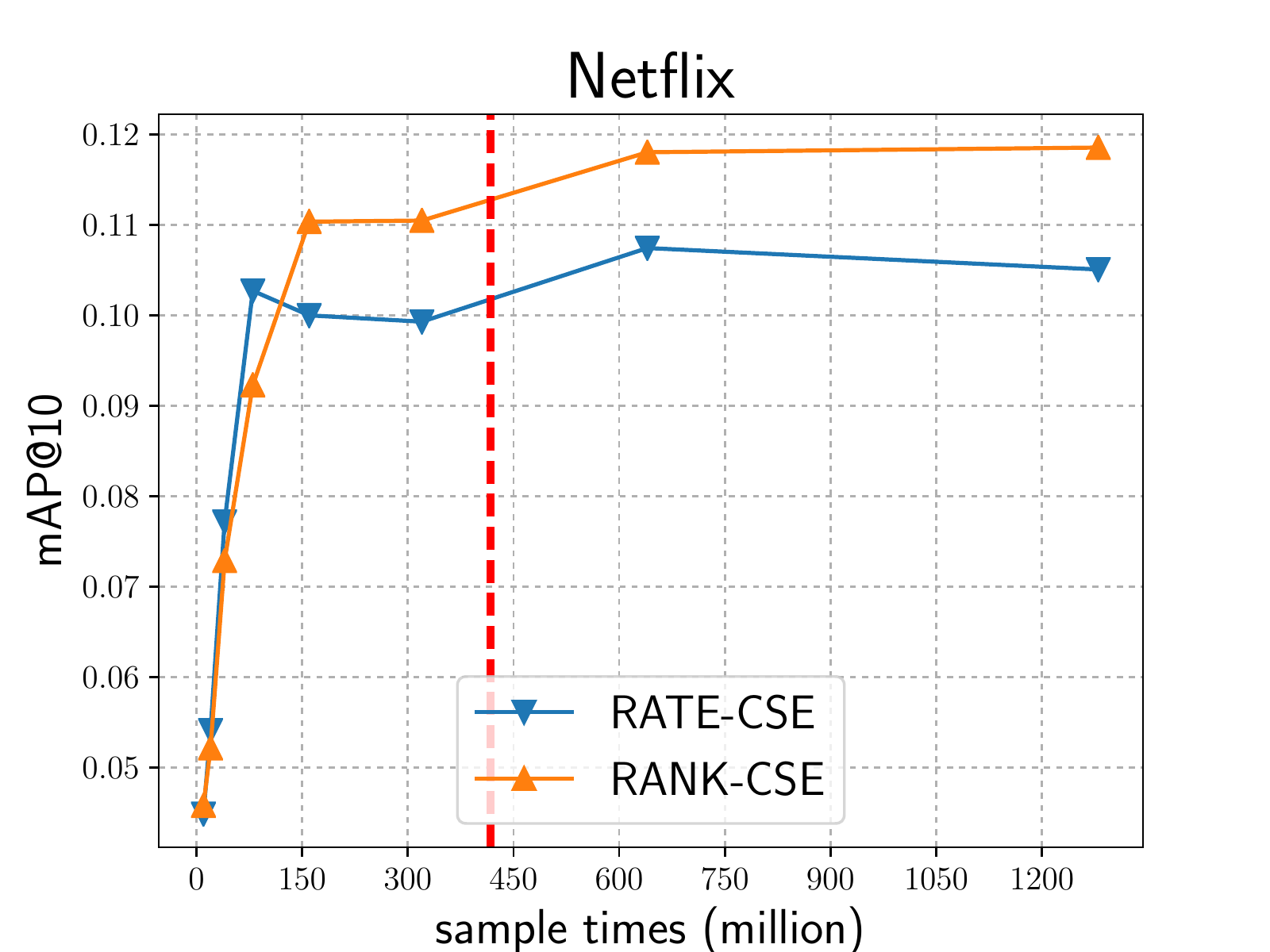}
\end{subfigure}
\hskip -2.5ex
\begin{subfigure}[b]{0.255\textwidth}
    \includegraphics[width=\textwidth,height=0.75\textwidth]{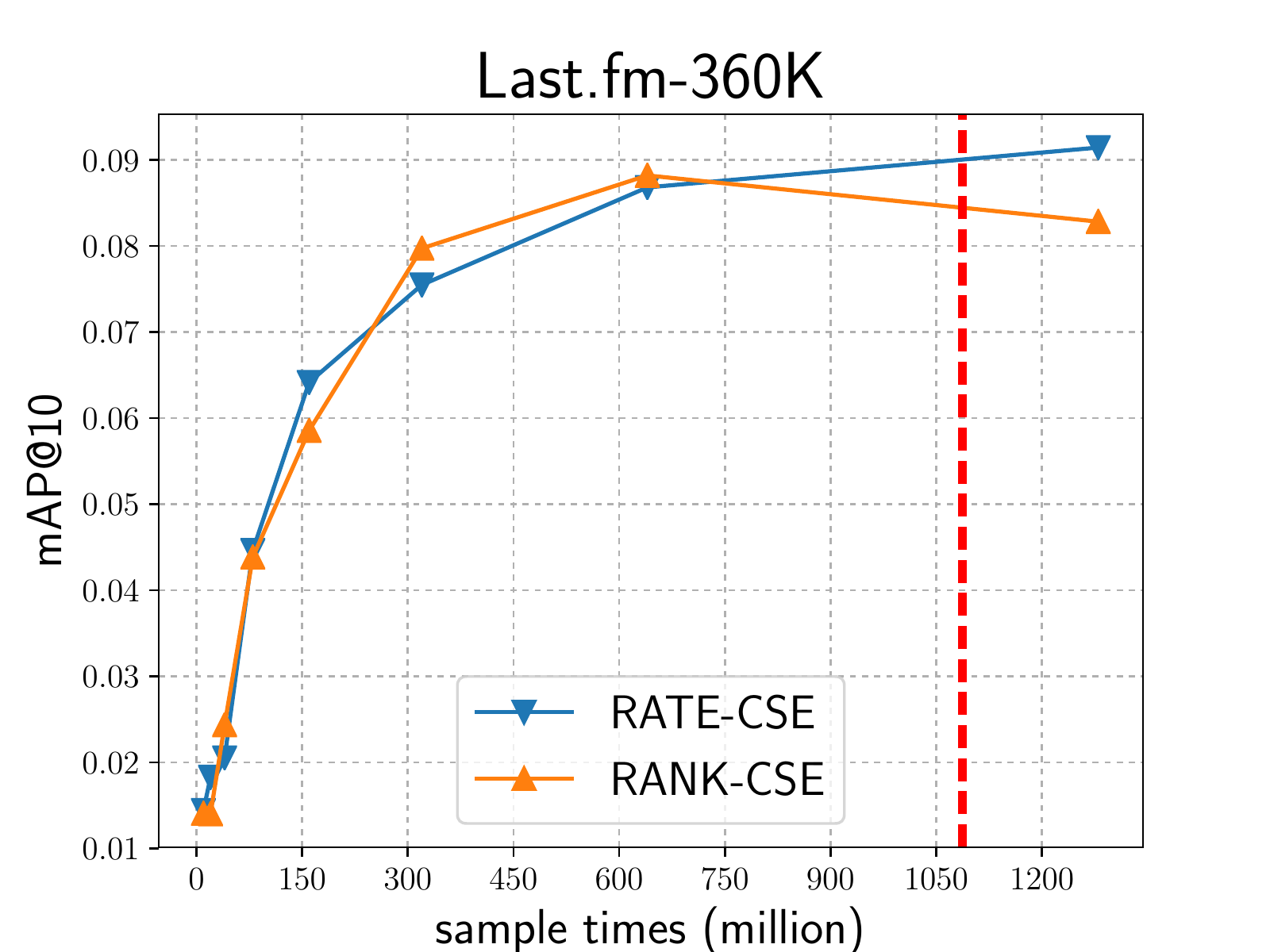}
\end{subfigure}
\hskip -2.5ex
\begin{subfigure}[b]{0.255\textwidth}
    \includegraphics[width=\textwidth,height=0.75\textwidth]{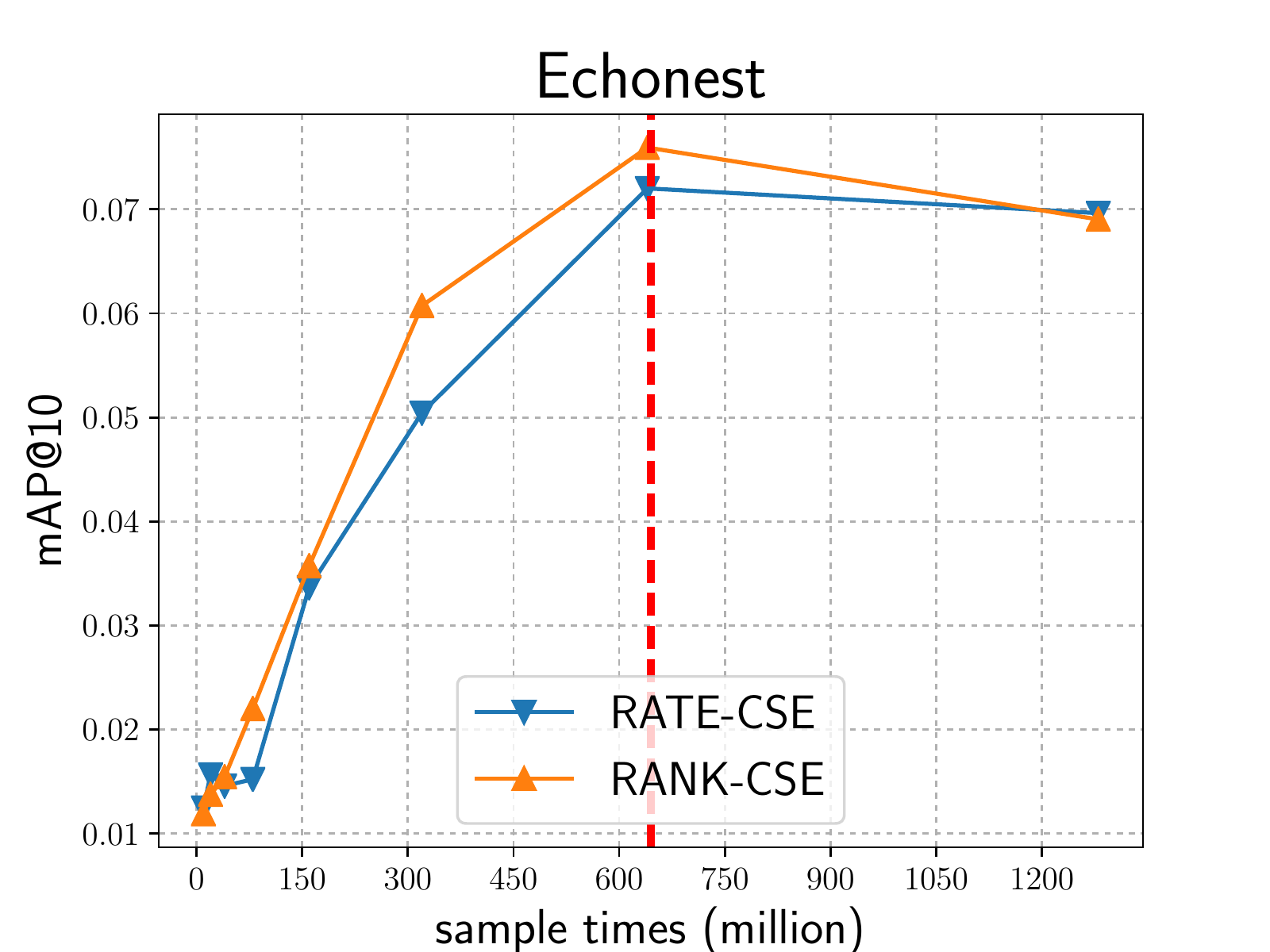}
\end{subfigure}

\caption{Sensitivity and convergence analyses} 
\label{fig:sen}
\end{figure*}

\subsubsection{Evaluations}
The performance is evaluated between the recommended list $R_u$ containing
top-$N$ recommended items and the corresponding ground truth list $T_u$ for
each user $u$. We consider following two commonly-used metrics over these $N$
recommended results:
\begin{itemize}
    \item Recall: Denoted as Recall@$N$, which describes the fraction of the
        ground truth (i.e., the user preferred items) that are successfully
        recommended by the recommendation algorithm:
        \begin{eqnarray}
            \text{Recall@}N = \sum_{u\in U} \frac{1}{|U|}
            \frac{\sum_{v\in R_u}\mathbbm{1}(v\in T_u)}{\min(N, |T_u|)}. \notag
        \end{eqnarray}
    \item Mean Average Precision: Denoted as mAP@$N$, computing the mean of the
      average precision at $k$ (AP@$k$) for each user $u$ as
        \begin{eqnarray}
            \text{mAP@}N &=& \frac{1}{|U|} \sum_{u\in U}
            \text{AP}_{u}@k
            \\ \notag
            &=& \frac{1}{|U|} \sum_{u\in U}
            \frac{\sum_{k=1}^{N} \text{P}_{u}(k) \times \mathbbm{1}(r_k\in T_u)}{\min(N,|T_u|)}, \notag
        \end{eqnarray}
\end{itemize}
where $r_k$ is the $k$-th recommended item and $\text{P}_{u}(k)$ denotes the
precision at $k$ for user $u$.
This is a rank-aware evaluation metric because it considers the positions of
each recommended item. 
For each dataset, the reported performance was averaged over 10 times; in each
time, we randomly split the data into 80\% training set and 20\% testing set.

\subsection{Results}

\subsubsection{Recommendation Performance Comparison}
The results for the ten baseline methods along with the proposed method are
listed in Table~\ref{tb:rec}, where RATE-CSE and RANK-CSE denote two versions
of our method that employ respectively
rating-based and ranking-based loss functions for user-item associations.
Note that the best results are always indicated by the bold font, and for
coFactor and BiNE we report only the experimental results on Frappe and
CiteULike because of resource limitations.\footnote{While the memory usage of
  coFactor implementation is $\mathcal{O}(|V|^2)$, BiNE's requires extensive
  computational time, e.g., more than 24 hours to learn the embedding for the
large dataset, Movielens-Latest.}
As discussed in Section~\ref{sec:ge}, DeepWalk is not suitable for user-item
recommendation as it make the users apart from items in the embedding space. 
In addition, observe that BiNE does not perform well in our experiments; such a
result is due to the fact that BiNE is a general network embedding model and
thus does not incorporate the regularizer in their objective function, which is
however an important factor for the robustness of recommendation performance.
Comparing the performance of the other baseline methods, we observe that
the performance of WALS, WARP and $K$-OS is very competitive. That is, these
methods achieve the top performance among all the baselines on several datasets.
The performance of WalkRanker and CML, on the other hand, seems
satisfactory only on two rather small datasets~-- Frappe and
CiteULike~-- and performs more poorly on most of the other datasets.

We observe that our method achieves the best results in terms of both Recall@10
and mAP@10 for most datasets. Moreover, RANK-CSE generally outperforms
RATE-CSE in the experiments, re-confirming that using a ranking-based loss is
indeed better for datasets with binary implicit feedbacks~\cite{bpr,warp,kos}.
Specifically, except for Frappe, RATE-CSE or RANK-CSE achieves significantly
much better performance than the best performing baseline methods with a
maximum improvement of +20.7\%.

\subsubsection{Parameter Sensitivity and Convergence Analyses}\label{sec:sens}

Figure~\ref{fig:sen} shows the results of the sensitivity analysis on two
hyper-parameters $k$ and $\lambda$ in the first and second rows, respectively, and
those of the convergence analysis based on sample times in the third
row.\footnote{Note that due to space limits, we report the results for the 
four largest datasets only.}
We first observe that increasing the order $k$ of modeling neighborhood proximity
between users or items improves the performance in general.
We first observe from Figure~\ref{fig:sen} is that the optimal value of $k$ is data
dependent and has to be empirically tuned considering the trade-off between
accuracy and time/space complexity.
In general, a larger $k$ leads to better result, and from our experience, the
result would reach a plateau when $k$ is sufficiently large (e.g., when $k>3$).
The second row of Figure~\ref{fig:sen} shows how the balancing parameter $\lambda$ affects
performance: RANK-CSE obtains better performance with a value around 0.05,
while RATE-CSE performs well with a value around 0.1. 
Finally, we empirically show that the required total sample times for 
convergence is linear with respect to $|E|$ as illustrated in the third row,
where the vertical dash line indicates the boundary of $|E|\times 80$ as
we applied to the previous recommendation experiment.
As the training time depends linearly on the sample times, it can be said
that both RATE-CSE and RANK-CSE converge with less than a constant multiple
of $|E|$ sample times. This demonstrates the nice scalability of CSE.

\section{Conclusion}\label{sec:conclude}

We present CSE, a unified representation learning framework that exploits
comprehensive collaborative relations available in a user-item bipartite graph
for recommender systems.
Two types of proximity relations are modeled by the proposed DSEmbed and
NSEmbed modules. Moreover, we propose a sampling technique to enhance the
scalability and flexibility of the model.
Experimental results show that CSE yields superior recommendation
performance over a wide range of datasets with different sizes, densities,
and types than many state-of-the-art recommendation methods.

\bibliographystyle{ACM-Reference-Format}
\bibliography{paper}

\end{document}